\documentclass[aps,twocolumn,prb,superscriptaddress]{revtex4}
\usepackage{graphicx}
\usepackage{dcolumn}
\usepackage{amsmath}
\usepackage{natbib}

\begin{document}

\title{Electromagnon and phonon excitations in multiferroic TbMnO$_3$}

\author{P. Rovillain}
\affiliation{Laboratoire Mat\'eriaux et Ph\'enom\`enes Quantiques UMR 7162 CNRS, Universit\'e Paris Diderot-Paris 7, 75205 Paris cedex 13, France}
\affiliation{School of Physics, University of New South Wales, Sydney, New South Wales 2052, Australia}
\affiliation{The Bragg Institute, ANSTO, Kirrawee DC NSW 2234, Australia}
\author{J. Liu}
\author{M. Cazayous}
\author{Y. Gallais}
\author{M-A. Measson}
\affiliation{Laboratoire Mat\'eriaux et Ph\'enom\`enes Quantiques UMR 7162 CNRS, Universit\'e Paris Diderot-Paris 7, 75205 Paris cedex 13, France}
\author{H. Sakata}
\affiliation{Department of Physics, Tokyo University of Science, 1-3 Kagurazaka Shinjyuku-ku Tokyo, Japan 162-8601}
\author{A. Sacuto}
\affiliation{Laboratoire Mat\'eriaux et Ph\'enom\`enes Quantiques UMR 7162 CNRS, Universit\'e Paris Diderot-Paris 7, 75205 Paris cedex 13, France}

\date{\today}
     
\begin{abstract}
We have performed Raman measurements on TbMnO$_3$ single crystal under magnetic field along the three crystallographic directions. The flip of the spin spiral plane creates an electromagnon excitation. In addition to the electromagnons induced by the Heisenberg coupling, we have detected the electromagnon created by the Dzyaloshinskii-Moriya interaction along the c axis. We have identified all the vibrational modes of TbMnO$_3$. Their temperature dependences show that only one phonon observed along the polarization axis is sensitive to the ferroelectric transition. This mode is tied to the Tb$^{3+}$ ion displacements that contribute to the ferroelectric polarization.   
\end{abstract}


\maketitle

\section{Introduction}

\par

Multiferroic materials exhibit simultaneously two or more ferroic orders, such as (anti)ferromagnetic, (anti)ferroelectric or (anti)ferroeleastic order. 
For some of these materials, the interaction between ferroelectricity and magnetism is especially strong and is of great interest for applications in future generations of novel storage or data processing devices, where magnetization can be controlled by an electric field and vice versa.\cite{Bea2008a, Eerenstein2006, Balke2012, Chanthbouala2012, Rovillain2010a}

In the so-called improper multiferroics such as perovskite manganites RMnO$_3$ (R rare earth), the electric polarization is induced by spin order breaking inversion symmetry via the spin-orbit interaction\cite{Hu2008}, exchange striction\cite{Mostovoy2006} or other related effects. TbMnO$_3$ is by far the most studied multiferroic manganite of this class. Such a compound is  fundamental to study novel coupling between microscopic degrees of freedom such as spin and lattice.\cite{Khomskii2009}

The static and dynamic properties are not necessarily induced by the same mechanisms. Symmetry analysis has been able to determine the spiral spin structure giving rise to an electric polarization\cite{Kenzelmann2005} but the question of the dominant microscopic mechanism is still unsolved. A purely electronic scenario based on the hybridization of electronic orbitals can explain the shift of the center of charge.\cite{Hu2008} A second mechanism involves a magnetically-induced ionic displacements resulting from the Dzyaloshinskii-Moriya interaction (DMI) proportional to \textbf{S$_n$} $\times$ \textbf{S$_m$}.\cite{Sergienko2006, Katsura2005} Recent measurements estimate the contribution of the ionic displacement around a quarter of the polarization value.\cite{Walker2011} Concerning the dynamic properties of multiferroics, these materials offer the opportunity to study novel excitations called electromagnons arising from the hybridization between spin waves and phonon modes. These magnetoelectric excitations are driven by the electric field of light.\cite{Pimenov2006} Electromagnons have been theoretically put forth by Smolenskii and Chupis\cite{Smolenskii1982} and their signatures have been recently evidenced by infrared (IR)\cite{Sushkov2007, Takahashi2008}, terahertz (THz)\cite{Pimenov2006} and Raman\cite{Rovillain2010} spectroscopies. Their energies match spin wave dispersion energies determined by inelastic neutron scattering.\cite{Senff2007} 

Considerable efforts have been devoted to determine the origin and symmetry of magnetoelectric coupling. Katsura et al.\cite{Katsura2007a} firstly proposed a model for the dynamical coupling based on the DMI developped to explain the ferroelectricity. In this model, the electromagnons should appear with \textbf{E$^{\omega}$} parallel to the ferroelectric polarization ($c$ axis in the $ab$ spiral spin ordered phase) in contradiction with IR and THz measurements ($a$ axis in all the spin ordered phase).\cite{Aguilar2009} A scenario based on the Heisenberg coupling mechanism between non-colinear spins can explain these optical selection rules for the electromagnons.\cite{Aguilar2009, Stenberg2009a} However, the exact spin spiral configuration of the ground state giving rise to electromagnons is still an open question.\cite{Stenberg2009a, Mochizuki2010}

In this article, we have investigated by Raman scattering the magnetoelectric phase diagram of TbMnO$_3$ along the three crystal directions. With a magnetic field $\textbf{B}$ along the $a$ axis we follow the frequency shift of two electromagnons. Along the $b$ axis an additional electromagnon excitation is induced when the spiral spin flips from the $bc$ to the $ab$ plane. 
Moreover we point out a low frequency mode only excited with the electric field of light parallel to the polarization. We have identified this excitation as the electromagnon mode induced by the DMI. We also present polarized Raman spectra that show all the A$_g$, B$_{1g}$, B$_{2g}$ and B$_{3g}$ phonon modes. The temperature dependence of the A$_g$ modes along the $c$ direction shows a renormalization of the 113 cm$^{-1}$ phonon mode at the ferroelectric transition. This mode related to Tb$^{3+}$ ions displacements
contributes to the ferroelectric polarization in agreement with recent X-ray measurements.\cite{Walker2011}

\section{Experimental Details}

\par
Single crystals of TbMnO$_3$ were grown by floating-zone method and aligned using Laue X-ray back-reflection. The crystals have been polished to obtain high surface quality for optical measurements. TbMnO$_3$ crystallizes in the orthorhombic symmetry (P\textit{bnm}) with lattice parameters equal to $a=5.3~\mathring{A}, b=5.86~\mathring{A}, c=7.49~\mathring{A} $.\cite{Alonso2000} TbMnO$_3$ becomes antiferromagnetic below the N\'eel temperature T$_N$=42~K.\cite{Quezel1977} In this phase the Mn magnetic moments form an incommensurate sinusoidal wave with a modulation vector along the $b$ axis. The ferroelectric order appears below T$_C$=28~K at the magnetic transition from incommensurate to commensurate order where the spin modulation becomes a cycloid.\cite{Kimura2003} In this phase the spin of Mn$^{3+}$ rotates in the $bc$ plan and the electric polarization appears along the $c$ axis. We have probed three TbMnO$_3$ single crystals with the $ab$ (xy), $ac$ (xz) and $bc$ (yz) planes. The measurements of the electromagnons have been performed using an electric field of incident and scattered light parallel to the crystallographic axis $a, b, c$ (e.i. x, y, z) at 10 K. The phonon modes have been obtained with parallel and cross polarizations as detailed below. 

\begin{figure}
 \includegraphics*[width=5cm]{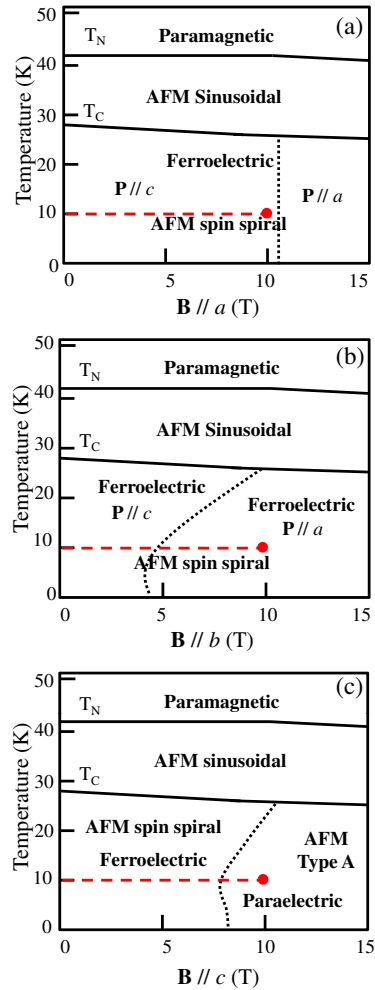} 
 \caption{\label{Fig1} 	 
 Magnetoelectric phase diagrams of TbMnO$_3$ for (a) \textbf{B}//$a$, (b) \textbf{B}//$b$ and (c) \textbf{B}//$c$ axis. Raman spectra have been recorded has a function of the magnetic field along the horizontal dashed paths.}
\end{figure}

Raman measurements are carried on in backscattering geometry using the 568 nm and 514.52 nm excitation lines from a Ar$^+$-Kr$^+$ mixed gas laser to investigate the electromagnon and phonon excitations, respectively. The scattered light is analysed by a triple spectrometer Jobin-Yvon T-64000 in soustractive configuration coupled to a liquid-nitrogen-cooled CCD detector. In this spectrometer configuration, the resolution of the phonon frequencies is less than 0.5 cm$^{-1}$.
The temperature dependence from 10 K to 300 K have been performed with a ARS closed-cycle He cryostat. The measurements under a magnetic field up to 10 T have been obtained using an Oxford Spectromag split-coil magnet.  

\section{Results and discussion}

Figure~\ref{Fig1} shows the magnetoelectric phase diagram of TbMnO$_3$ with magnetic field \textbf{B} along the (a) $a$, (b) $b$ and (c) $c$ axis.\cite{Kimura2005, Argyriou2007}
The diagrams show at zero Tesla the high temperature paramagnetic phase, the incommensurate sinusoidal phase below the N\'eel temperature T$_N$=42~K and the ferroelectric phase at T$_C$=28~K  where the spin modulation becomes a cycloid.\cite{Quezel1977, Kimura2003}

The magnetic field along the $a$ axis induces the flip of cycloid plane from the $bc$ to the $ab$ plane at high value above 10 Tesla (doted line in Fig.~\ref{Fig1}(a)). 
This value is reduced with a magnetic field along the $b$ axis down to 4-5 T at low temperatures (Fig.~\ref{Fig1}(b)).  
In Fig.~\ref{Fig1}(c) the effect of a magnetic field along the $c$ axis is distinct from along $a$ and $b$ axis. The cycloid structure is destabilized starting at 7 T. TbMnO$_3$ becomes paraelectric above 8 T and paramagnetic above 10 T. 

\subsection{Electromagnons}

\begin{figure}
 \includegraphics*[width=8.5cm]{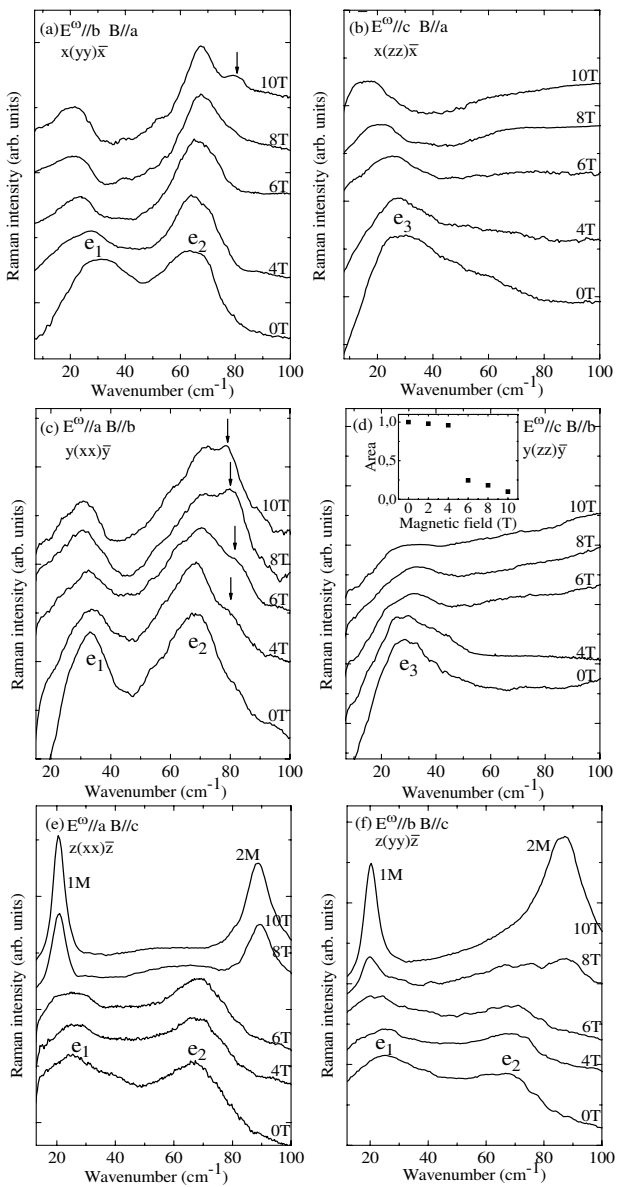} 
 \caption{\label{Fig2} 	 
Raman spectra measured using different configuration for the electric \textbf{E$^{\omega}$} and magnetic \textbf{H$^{\omega}$} fields of light on $(bc)$, $(ac)$ and $(ab)$ samples with a magnetic field \textbf{B}//$a$; $b$ and $c$ axis, respectively. The measurements have been performed at 10 K along the paths drawn in the magnetoelectric phase diagram of Fig.~\ref{Fig1}. e$_1$, e$_2$ and e$_3$ correspond to electromagnon modes. 1M and 2M correspond to the zone center magnon and to the two magnon mode of the zone edge, respectively. Arrows indicate the additional electromagnon resonance induced by the magnetic field modulation of the spin structure.}
\end{figure}

The electromagnons are magnetic excitations with an electric dipole. In IR and THz spectroscopies, they are activated by the electric field of light \textbf{E$^{\omega}$} whereas conventional magnetic excitations are driven by the magnetic field component \textbf{H$^{\omega}$} and appears in the magnetic permeability spectrum. Pimenov et al.\cite{Pimenov2006} have observed this new class of excitations in RMnO$_3$ multiferroics as magnetic resonances in the dielectric constant spectrum. 
In TbMnO$_3$, two electromagnons e$_1$ and e$_2$ are measured, respectively, at around  25 cm$^{-1}$ (3.2 meV) and 60 cm$^{-1}$ (7.4 meV) using IR\cite{Aguilar2009}, THz\cite{Takahashi2008} and Raman\cite{Rovillain2010} spectroscopies (Fig.~\ref{Fig2}(a) at 0 T).
IR and THz spectroscopies reveal that these absorptions are only electrically active for \textbf{E$^{\omega}$}//$a$ axis regardless of the cycloid plane orientation.\cite{Aguilar2009} Polarized inelastic neutron scattering experiments on TbMnO$_3$ have suggested that the e$_1$ mode is a magnetic zone center excitation.\cite{Senff2008a}
The e$_2$ mode is associated with the zone edge excitation. Both excitations correspond to propagating modes of the spins out of the cycloid plane. Recent spectrocopic studies have clarified that the wave vectors of the e$_1$ and e$_2$ electromagnons are equal to q$_{e_1}$ = $\pi$-2Q$_b$ (Q$_b$ is the magnitude of the cycloidal wave vector) and q$_{e_2}$ = $\pi$.\cite{Rovillain2011}

The dynamical magnetoelectric effect proposed by Katsura et al.\cite{Katsura2007a} on the basis of their model for the ferroelectricity failed to explain the IR and THz optical selection rules. 
The DMI could not thus only explain the origin of the both e$_1$ and e$_2$ electromagnons. It turned out in Ref.~\onlinecite{Aguilar2009} that the e$_2$ electromagnon originates from the conventional magnetostriction mechanism \textbf{S$_n$} . \textbf{S$_m$} whereas the origin of the lower lying electromagnon e$_1$ is still an open question. In addition to the magnetoelectric coupling, the spin orbit interaction\cite{Stenberg2009a} or the cycloid anharmonicity\cite{Mochizuki2010} have been proposed for its origin. Notice that an additional electromagnon activated with \textbf{E$^{\omega}$} parallel to the $c$ axis has been observed in THz spectroscopy around 20 cm$^{-1}$ (2.48 meV) as predicted by the model based on the DMI.\cite{Aguilar2009, Shuvaev2010}
Figure~\ref{Fig2} shows the Raman spectra measured on the $bc$, $ac$ and $ab$ planes under a magnetic field along the $a$, $b$ and $c$ axis, respectively.
At 0 T, Fig.~\ref{Fig2}(a) shows the both e$_1$ and e$_2$ electromagnons activated with an electric field of light \textbf{E$^{\omega}$}//$b$ axis on the $bc$ samples (x(yy)\={x} scattering geometry). 
Unlike IR and THz spectroscopies that observe electromagnon only with \textbf{E$^{\omega}$}//$a$, the e$_1$ and e$_2$ electromagnons are observed using a polarization of the electric field of light along the $a$ and $b$ axis (Fig.~\ref{Fig2}(a) and (c)). This difference points out the specific selection rules of the Raman spectroscopy. 
In conventional Raman spectroscopy, \textbf{H$^{\omega}$} does not couple directly to spin excitations. Instead \textbf{E$^{\omega}$} couples to magnetic excitations via the spin-orbit coupling.\cite{Fleury1968a} The observation of electromagnons using Raman scattering is thus indirect in comparison with the other optical techniques.  

Several arguments have been developed in Ref.~\onlinecite{Rovillain2010} and in Ref.~\onlinecite{Rovillain2011} to associate the Raman modes e$_1$ and e$_2$ to electromagnons excitations. Both peaks disappear at the Curie temperature well below the N\'eel temperature (a pure magnetic excitation usually desapears around the the N\'eel temperature). The width of these peaks is broader than the pure magnon mode and they appear at different energies (see Fig.~\ref{Fig2}(e)). The energies of the e$_1$ and e$_2$ modes correspond to the energies of the broad peaks already reported by IR around 25 cm$^{-1}$ and 60 cm$^{-1}$  and assigned to electromagnon excitations. The peak at 60 cm$^{-1}$ has an energy close to the magnetic zone-edge energy. Raman scattering probes the dispersion branches close to the zero wave vector. The activation of 60 cm$^{-1}$ zone-edge magnon could be explained by the alternation of the Heisenberg exchange interaction along the $b$ axis\cite{Aguilar2009} or by the coupling of this mode with the spontaneous polarization through the dynamical magnetoelectric field\cite{Stenberg2009a}. The peak at 30 cm$^{-1}$ might be assigned to zone-center magnon mode. However the Raman signature of pure zone-center magnon mode is clearly different (both in energy and width) (see Fig ~\ref{Fig2}(e)) indicating a significant coupling to ferroelectric degrees of freedom.

In Fig.~\ref{Fig2}(a) the increase of the magnetic field along the $a$ axis induces the down shift of the e$_1$ peak and the up shift of the e$_2$ peak. 
At high magnetic field (10 T), a new excitation appears around 80 cm$^{-1}$. Above 10 T, the spin spiral is expected to flip from the $bc$ to the $ab$ plane (Fig.~\ref{Fig1}(a)). In Ref.~\onlinecite{Rovillain2011} we have compared our measurements to the calculated spectral weight of electromagnons under a magnetic field along the $b$ axis in Heisenberg model. In the calculations, a peak shows up clearly around 90 cm$^{-1}$ for B=8 T when the spin spiral flips in the ab plane. This additional peak comes from the exchange and easy-plane anisotropy terms and has been interpreted as an electromagnon resonance.\cite{Mochizuki2010, Rovillain2011}

In Fig.~\ref{Fig2}(b) Raman spectra obtained using \textbf{E$^{\omega}$}//$c$ (x(zz)\={x} configuration) exhibit only one peak e$_3$ $\approx$ 30 cm$^{-1}$ at the energy of the e$_1$ excitation. Increasing the magnetic field the 
e$_3$ peak frequency decreases. The electromagnon origin of the e$_3$ mode is supported by the same arguments used for the e$_1$ mode. They distinguish each other by different Raman selection rules. Unlike the e$_1$ and e$_2$ electromagnons we will observe that this excitation is tied to polarization direction. As mentioned before, only the electromagnons induced by the DMI have been observed with \textbf{E$^{\omega}$} parallel to the polarization axis. The activation of an electromagnon originating from the DMI is further discussed below.

Figure \ref{Fig2}(c) and (d) present Raman spectra recorded on the $ac$ samples that allow an investigation of the magnetoelectric phase diagram with $\textbf{B}$//$b$.
In Fig.~\ref{Fig2}(c), the frequency shift of the e$_1$ and e$_2$ peaks is similar to the one measured in Fig.~\ref{Fig2}(a) with a magnetic field along $a$. The additional peak at 80 cm$^{-1}$ associated with an electromagnon excitation appears at lower magnetic field (around 4 T) compared to Fig.~\ref{Fig2}(a) as expected from the magnetoelectric phase diagram in Fig.~\ref{Fig1}(b) where the flip of the spin spiral plane occurs between 4 and 5 T. The e$_3$ peak is again observed using \textbf{E$^{\omega}$}//$c$ (y(zz)\={y} scattering geometry) in  Fig.~\ref{Fig2}(d). However, the magnetic field dependence of e$_3$ is clearly different compared to its behavior with a magnetic field along $a$. Here, the intensity of the e$_3$ peak is constant up to 4 T and decreases steeply at 6 T. The normalized area of the peak as a function of the magnetic field reported in the inset of Fig.~\ref{Fig2}(d) shows quantitatively this behavior. Around 10 T, the peak has almost vanished. Remember that between 4 and 5 T, the spin spiral plane flips from the $bc$ to the $ab$ plane with the simultaneous flip of the spontaneous polarization from the $c$ to the $a$ axis. An electromagnon induced by the DMI is tied to the polarization axis and should be no more activated with \textbf{E$^{\omega}$}//$c$ (i.e. x(zz)\=x or y(zz)\=y) when the polarization flips along the $a$ axis. The e$_3$ peak follows this behavior. The peak is activated when \textbf{E$^{\omega}$} is parallel to the polarization axis and its intensity strongly decreases when the polarization is perpendicular to \textbf{E$^{\omega}$}. This argument is in favor of the interpretation of the e$_3$ peak as the signature of the electromagnon due to the DMI and confirm the previous observation of a weak resonance in THz measurements.\cite{Shuvaev2010} Notice that, when the polarization flips from the $c$ to the $a$ axis, the e$_3$ peak should be activated with \textbf{E$^{\omega}$}//$a$ (y(xx)\=y) and should appear in Fig.~\ref{Fig2}(c) above 4 T. However, the e$_1$ peak is located at the same  energy and prevents a clear observation of the e$_3$ peak. 

\begin{figure}
 \includegraphics*[width=8cm]{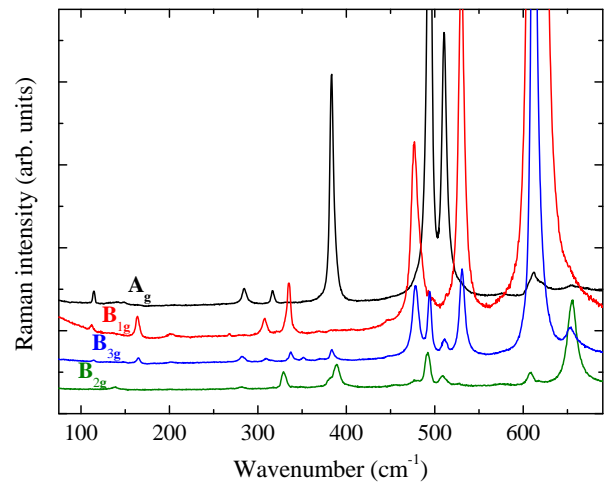} 
 \caption{\label{fig1} 	 
 Polarized Raman spectra of TbMnO$_3$ phonon modes at 10 K. The A$_g$ modes (black curve) are observed with parallel polarization of the incident and scattered light. The $B_{1g}, B_{2g}$ and $B_{3g}$ modes are observed for a crossed polarization (xy), (xz) and (yz), respectively.}
\end{figure}

Figures \ref{Fig2}(e) and (f) show the Raman responses with a polarization of the light \textbf{E$^{\omega}$} parallel to the $a$ and $b$ axis measured on the $ab$ sample (i.e. z(xx)\=z and z(yy)\=z configurations, respectively) under a magnetic field $\textbf{B}$//$c$. 
In the both figures, the frequencies of the e$_1$ and e$_2$ electromagnons decrease and increase with the magnetic field respectively. Around 8 T, one sharp peak 1M appears at 21 cm$^{-1}$ and a broader one 2M at 88 cm$^{-1}$. The intensity of the electromagnons is reduced and disappears at 10 T in Fig.~\ref{Fig2}(e), and in Fig.~\ref{Fig2}(f). The 1M peak is associated to the pure magnon excitation at the zone center of the spin wave dispersion in the spin spiral state. The 2M peak corresponds to the two magnon excitations resulting from two zone-edge magnons with opposite wave vectors. The energy of these peaks is in agreement with neutron measurements.\cite{Senff2008a}
Figures \ref{Fig2}(e) and (f) point out the spectral weight transfer from the electromagnons to the magnon excitations entering in the paraelectric phase and illustrate the dehybridization of the electromagnons with a magnetic field $\textbf{B}$//$c$.\cite{Rovillain2011} 

\subsection{Phonon modes}

\begin{table}[ht]
  \begin{center}
  \parbox{8cm}
  \caption{
   
  Frequency (cm$^{-1}$) of phonon modes in TbMnO$_3$ single crystal. The assigment of the modes to atomic motions is reported in Ref.~\citenum{Iliev1998}}\\
	\vspace{3mm}
	 \begin{tabular}{ccccccc}
	  \multicolumn{2}{c}{$\Gamma$-point} & \multicolumn{3}{c}{Experimental} & \multicolumn{2}{c}{Calculations} \\
	\hline\noalign{\smallskip}
	     & & Rovillain & Laverdiere\cite{Laverdiere2006a} & Iliev\cite{Iliev2006} & Gupta\cite{Gupta2008a} & Choitrani\cite{Choithrani2011} \\
	      & & 10K & 10K & 300K &  &  \\
	\hline\noalign{\smallskip}
	  A$_g$ & (5) & 113 & -   & -   & 79  & 108 \\
	        & (6) & 146 & -   & -   & 211 & 146 \\
	        & (2)  & 283 & 284 & 280 & 269 & 236 \\
	        & (7)  & 315 & 316 & 314 & 378 & 273 \\
	        & (4)  & 381 & 384 & 378 & 402 & 338 \\
	        & (1)  & 492 & 494 & 488 & 489 & 446 \\
	        & (3)  & 509 & 511 & 509 & 509 & 510 \\
	  \hline\noalign{\smallskip}
	  B$_{1g}$ & (6) & 165 & -   & -   & 96  & 100 \\
	  				 & (4) & 309 & 309 & -   & 127 & 199 \\
	  				 & (7)  & 335 & 336 & 330 & 331 & 287 \\
	  				 & (5)  & 369 & -   & -   & 474 & 350 \\
	  				 & (3)  & 477 & 478 & 473 & 501 & 498 \\
	  				 & (2)  & 530 & 531 & 528 & 528 & 524 \\
	  				 & (1)  & 613 & 614 & 612 & 612 & 573 \\
	  \hline\noalign{\smallskip}
	  B$_{2g}$ & (5) & 140 & - & - & 134 & 132 \\
	           & (4) & 280 & - & - & 302 & 206 \\
	           & (3) & 331 & - & - & 469 & 268 \\
	           & (2) & 389 & - & - & 519 & 296 \\
	           & (1) & 608 & - & - & 621 & 530 \\
	  \hline\noalign{\smallskip}
	  B$_{3g}$ & (5) & 135 & - & - & 127 & 102 \\
	           & (4)  & 309 & - & - & 270 & 249 \\
	           & (3)  & 337 & - & - & 381 & 284 \\
	           & (2)  & 351 & - & - & 432 & 454 \\
	           & (1)  & 384 & - & - & 545 & 504 \\
	  \hline\noalign{\smallskip}
	 \end{tabular}
	\end{center}
	\label{freqphonon}
 \end{table}

TbMnO$_3$ is an orthorhombic distorted perovskite structure with the space group $Pbnm$. The vibrational modes detected by Raman spectroscopy depend on   
the crystal symmetry which controls the matrix elements of  the Raman tensor and on  
the incident and scattered light polarizations which stress the Raman tensor. 
The irreducible representation for the normal modes in this system gives 60 phonon modes at the $\Gamma$-point: $7A_g+7A_u+7B_{1g}+10B_{1u}+5B_{2g}+8B_{2u}+5B_{3g}+10B_{3u}$ and 24 of these modes are Raman allowed: $\Gamma_{Raman} = 7A_g+7B_{1g}+5B_{2g}+5B_{3g}$.
All the measurements have been performed in backscattering configuration 
(incident wave vector parallel to the scattered one). In order to get the pure A$_g$ modes we have used 
parallel polarizations corresponding to $y(zz)\bar{y}$ geometry.\cite{Porto1966} 
The $B_{1g}, B_{2g}$ and $B_{3g}$ modes have been obtained from crossed polarizations $z(xy)\bar{z}$, $y(xz)\bar{y}$ and $x(yz)\bar{x}$, respectively.\\

\par
Figure \ref{fig1} shows the Raman spectra of TbMnO$_3$ at 10 K in parallel (zz) and crossed (xy), (xz) and (yz) polarizations. We have measured all the 7~A$_g$ phonon modes in parallel polarization and the 7~B$_{1g}$, 5~B$_{2g}$ and 5~B$_{3g}$ phonon modes in crossed polarization as expected by the group theory.
The frequencies of the phonon modes at 10~K are reported in Table I and compared to the previous experimental results of J. Laverdi\`ere \textit{et al.}\cite{Laverdiere2006a} at 5~K and M. N. Iliev \textit{et al.}\cite{Iliev2006} at room temperature. The theoretical calculations of H. C. Gupta \textit{et al.}\cite{Gupta2008a} and R. Choithrani \textit{et al.}\cite{Choithrani2011} are also reporter in Table I. In addition to the preview experimental results, we have measured the phonon modes at low frequency below 200 cm$^{-1}$. These low frequency phonon modes correspond to the vibrations of the rare earth Tb$^{+3}$ ions\cite{Venugopalan1985, Martin-Carron2002}. Moreover, we have measured the unreported B$_{2g}$ and B$_{3g}$ vibrational modes.\\

\begin{figure}
 \includegraphics*[width=4.37cm]{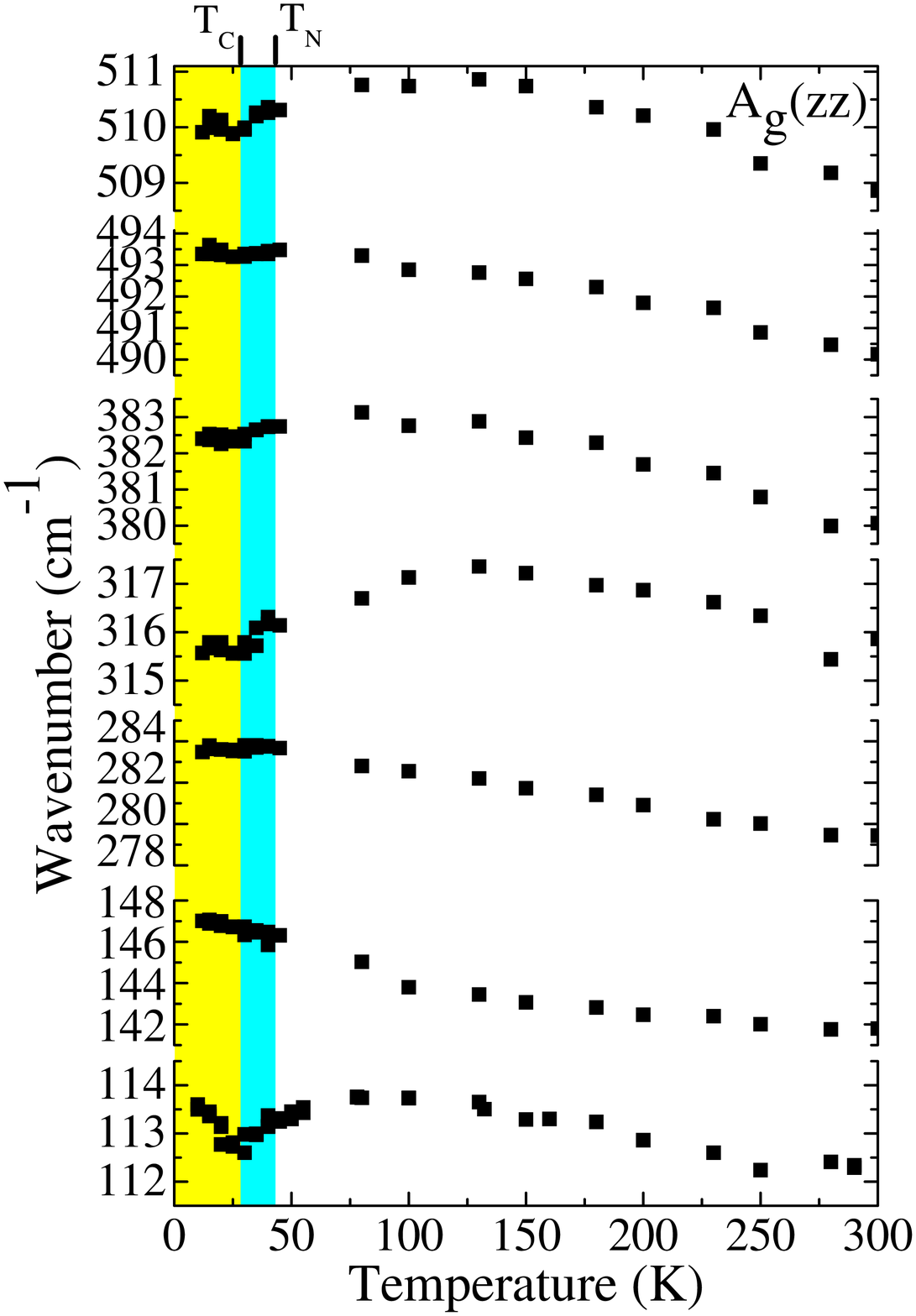} 
 \includegraphics*[width=4.03cm]{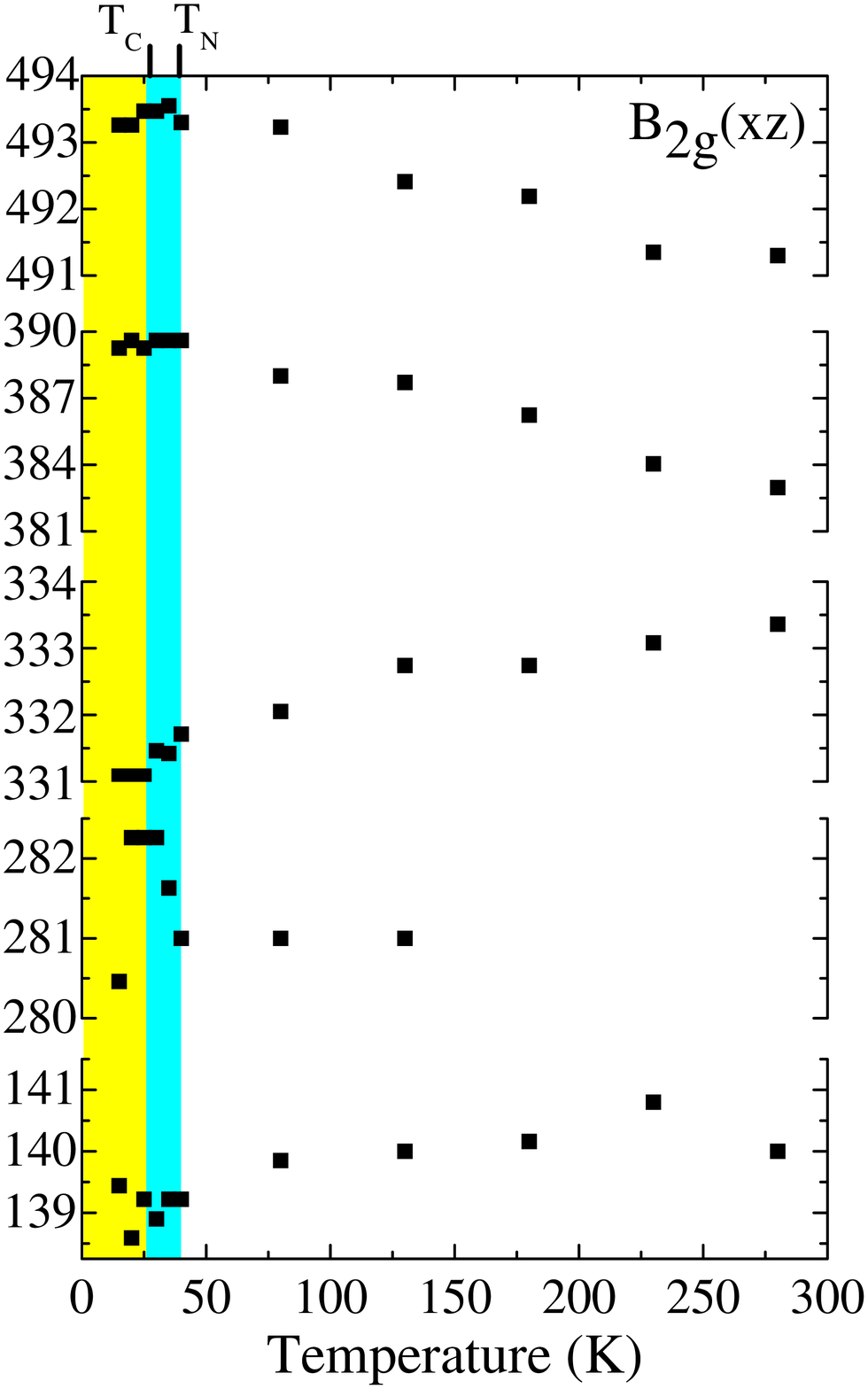}
 \caption{\label{fig2} 	 
 Temperature dependance of the A$_g$ and B$_{2g}$ phonon frequencies.}
\end{figure}

In order to analyze the effects of the phase transitions on the phonon modes, one can follow the temperature dependences. The phonon frequency usually tend to soften due to
the dilation of the unit cell when temperature increases. Except for three modes, all frequencies are higher at low temperatures.
The A$_g$(7) mode frequency increases up to 125 K and after decreases. This mode corresponds to the in-plane phase MnO$_6$ rotations around the $y$ axis. The B$_{2g}$(3) mode presents a continuous 
decrease of its frequency from high to low temperature. This mode is related to the out-of-phase MnO$_6$ bending. Such behaviors might be related to spin phonon coupling. 
IR measurements have shown an additional renormalisation of the phonon mode along the $c$ direction at 185 cm$^{-1}$ involving displacement of Mn at the ferroelectric transition.\cite{Schleck2010} Our Raman measurements point out the unusual behavior of the MnO$_6$ block frequencies probably tied to spin phonon interaction but not directly connected to the magnetic or ferroelectric transition. 

One question remains. Can we measure a soft mode associated with the ferroelectric transition that supports a mechanism involving a ionic displacement?
The temperature dependance of the A$_g$ mode frequencies have been measured in the (zz) polarization. The electric field wave vector of the incident and scattered photons are parallel to the $c$ axis
and thus parallel to the spontaneous polarization. 

The frequency of the mode at 113 cm$^{-1}$ in Fig.~\ref{fig2} presents a softening at the ferroelectric transition followed by a hardening. This partial softening reveals a mode softening behavior at the ferroelectric phase transition. The mode at 113 cm$^{-1}$ has been identified as related to the vibration of the Tb$^{3+}$ ions along the $c$ axis.\cite{Venugopalan1985,Martin-Carron2002} Recent improved X-ray diffraction technique has been able to measure the displacement along the $c$ axis of the Tb$^{3+}$ ions.\cite{Walker2011} These measurements established that this displacement contributes to a quarter of the electrical polarization magnitude. The softening of the 113 cm$^{-1}$ mode can be associated to the Tb$^{3+}$ ions displacement along the $c$ axis that contributes to the electrical polarization. Pure electronic scenario could be at the origin of the additional shift of the center of charge. 

Investigations of the temperature dependance of the Raman phonon modes should reveal the coupling of phonon and spin wave in particular to measure the spectral weight transfer from the phonon modes to the electromagnons. IR experiments have shown that the polar activity of the electromagnons comes from phonon even if there is no consensus on the phonon modes involved (phonon at 120 cm$^{-1}$ in Ref.~\citenum{Takahashi2008}, 133 cm$^{-1}$ in Ref.~\citenum{Schmidt2009} and 185 cm$^{-1}$ in Ref.~\citenum{Schleck2010} ). No spectral weight transfer have been detected in our temperature dependences. Previous Raman measurements under magnetic field along the $c$ axis in Ref.~\citenum{Rovillain2011} have shown that the A$_g$(2) and B$_{1g}$(1) modes both related to the Mn-O bound are involved in the polar activity of the electromagnons. 

\section{conclusion}
In summary, our measurements show the behavior of the electromagnons over all the magnetoelectric phase diagram. Two additional electromagnons have been detected. One is due to the spin structure modulation induced by the magnetic field, the other one corresponds to the electromagnon created by the DMI along the c axis. The investigation of the phonon temperature dependance points out the role of the phonon mode of Tb$^{3+}$ ions to the ferroelectric polarization.

\section*{Acknowledgments}
This work was supported in part by the Australian Research council (grant DP-110105346).

\end{document}